\documentclass{ws-ijmpa}
\usepackage[super,compress]{cite}
\usepackage{graphicx}
\usepackage{xspace}
\usepackage{url}
\usepackage{hyperref}
\hypersetup{colorlinks,citecolor=black,filecolor=black,linkcolor=black,urlcolor=blue}

\usepackage[table,xcdraw]{xcolor}
\usepackage[normalem]{ulem}
\usepackage{soul}

\usepackage{lineno}

\usepackage{orcidlink}
\newcommand{\orcidA}{\orcidlink{0000-0003-2849-0120}} 
\newcommand{\orcidB}{\orcidlink{0000-0001-9223-6480}} 
\newcommand{\orcidC}{\orcidlink{0000-0001-5038-678X}} 

\begin{document}
\markboth{G. Bíró, G. Papp, G. G. Barnaföldi}{Estimating event-by-event multiplicity by a Machine Learning Method for Hadronization Studies}

\title{Estimating event-by-event multiplicity by a Machine Learning Method \\ for Hadronization Studies}

%
\catchline{}{}{}{}{}
%

\author{Gábor Bíró\orcidA$^{\ast,\dagger}$}
\address{biro.gabor@wigner.hun-ren.hu}

\author{Gábor Papp\orcidC$^{\dagger}$}
\address{pg@elte.hu}

\author{Gergely Gábor Barnaföldi\orcidB$^{\ast}$}
\address{barnafoldi.gergely@wigner.hun-ren.hu}

\address{$^{\ast}$HUN-REN Wigner Research Centre for Physics\\29--33 Konkoly--Thege Mikl\'os \'ut\\ H-1121 Budapest, Hungary \\ 
$^{\dagger}$E\"otv\"os Lor\'and University, Institute of Physics and Astronomy\\1/A P\'azm\'any P\'eter S\'et\'any\\ H-1117  Budapest, Hungary\\}

\maketitle


\begin{abstract}
Hadronization is a non-perturbative process, which theoretical description can not be deduced from first principles. Modeling hadron formation requires several assumptions and various phenomenological approaches. Utilizing state-of-the-art Deep Learning algorithms, it is eventually possible to train neural networks to learn non-linear and non-perturbative features of the physical  processes. In this study, the prediction results of three trained ResNet networks are presented, by investigating charged particle multiplicities at event-by-event level. The widely used Lund string fragmentation model is applied as a training-baseline at $\sqrt{s}= 7$~TeV proton-proton collisions. We found that neural-networks with $ \gtrsim\mathcal{O}(10^3)$ parameters can predict the event-by-event charged hadron multiplicity values up to $ N_\mathrm{ch}\lesssim 90 $.


\end{abstract}

\ccode{PACS numbers: 13.85.-t, 12.38.-t, 02.70.-c, 02.70.Uu}


\maketitle


\section{Introduction}
\label{sec:introduction}

Color confinement is one of the most intriguing aspects of Quantum Chromodynamics (QCD). Due to the running nature of the strong coupling, the cross section of the partonic level scatterings can be well calculated with perturbation techniques at high energy scales. In contrary, hadronization --- the confinement of partons into hadrons --- occurs at lower energies, where calculations can not be performed perturbatively from first principles. Investigating this non-perturbative regime requires phenomenological models, where QCD-related scaling is well hided due to the high level of non-linearity~\cite{Bjorken:1968dy}. 

Machine Learning (ML) algorithms are able to describe non-perturbative (non-linear) processes in high-energy physics~\cite{Feickert:2021ajf}. This raises the question whether such ML methods could provide new solutions or generation tools in these soft regions, where traditionally complex Monte Carlo algorithms are used to describe hadron formation. It would be also interesting to identify scaling patterns as well. Therefore, the main goal of this study is the investigation of hadronization, by applying popular deep learning (DL) algorithms. Our primary physical motivation here is to develop a novel model of hadronization, while in parallel our aim is to extract correlations via the evaluation of the properties of the trained Deep Neural Networks (DNN). 

An application-oriented feature of this investigation can have a further beneficial impact on the future data analysis tasks. During the Long Shutdown 2 (LS2) of the Large Hadron Collider (LHC), all of the large detectors went through a major upgrade~\cite{Evans:2008zzb}. By approaching the High Luminosity era, it is expected to record an enormous amount of raw data, more than 200 PB each year. This is almost the quantity that was collected during Run~1 and~2 altogether. Additionally, the amount of the simulated data has to be increased along with the real experimental data, which is already a challenging task: even for a state-of-the-art hardware, to simulate only one second of LHC data, the necessary computing time is larger by several orders of magnitude. As a solution for this ever-growing difficulty, one can aim towards several directions:
\begin{enumerate}
  \item Development of new-generation algorithms and simulation softwares with built-in parallelization and/or hardware optimization~\cite{Grindhammer:1989zg, Biro:2019ijx, Amadio:2020ink, Butter:2020abv};
  \item Utilization of novel Machine Learning approaches to accelerate the numerical calculations~\cite{Monk:2018zsb, Vallecorsa_2018, Butter:2020abv}.
\end{enumerate}
In the sense of computing efficiency, the high potential of the various ML techniques already has been recognized in numerous fields. Beside studying new physics and theories, the development of accelerated numerical calculations also motivates our investigations. In the current work, we calculate the event-by-event final-state multiplicities of high-energy proton-proton collisions with DNNs.

\section{Brief overview of earlier related studies}
\label{sec:related}

Recently, the number of high-energy physics related applications of machine learning techniques has been increasing rapidly. Ref.~\cite{Feickert:2021ajf} is a dynamically growing collection of the latest publications related to the field. Among the various topics, a few results serve as a prequel to the current study.

A popular approach in the HEP community is to consider some aspects of a high-energy collisional event as a pixelated image, where e.g. the energy of the particles are collected in discretized pixels, spreading in the azimuth-polar plane. This approach resembles to the granularity of the detector components and very suitable for convolutional networks. The range of possible applications is very wide, spreading from (and not limited to) jet tagging~\cite{Cogan:2014oua}, parton showering~\cite{Barnard:2016qma, Monk:2018zsb} or estimating collective effects~\cite{Mallick:2022alr}. Other studies aim to identify jets and their properties~\cite{Moreno:2019bmu, Du:2021pqa, Du:2021brx, Dreyer:2021hhr}. In Ref.~\cite{Nguyen:2018ugw} the authors implement an event selection and classification algorithm based on the event images, by investigating convolutional architectures.

Several other methods and novel techniques are being developed to handle data in a more specialized approach as well, which are out of the scope of the current study. The utilization of point clouds as data structures is a natural way to process unordered, permutation invariant data with variable event sizes~\cite{Mikuni:2021pou, Qu:2019gqs}. Graph neural networks on the other hand provide an excellent tool to represent e.q. QCD processes or geometrical structures, where the pairwise relations are important~\cite{Shlomi:2020gdn}. Many other related works are collected and summarized in the HepML webpage~\cite{Feickert:2021ajf}.

\section{General structure of high-energy hadron-hadron collisions}
\label{sec:MCsims}

\begin{figure}[h]
\centering
\includegraphics[width=0.85\linewidth]{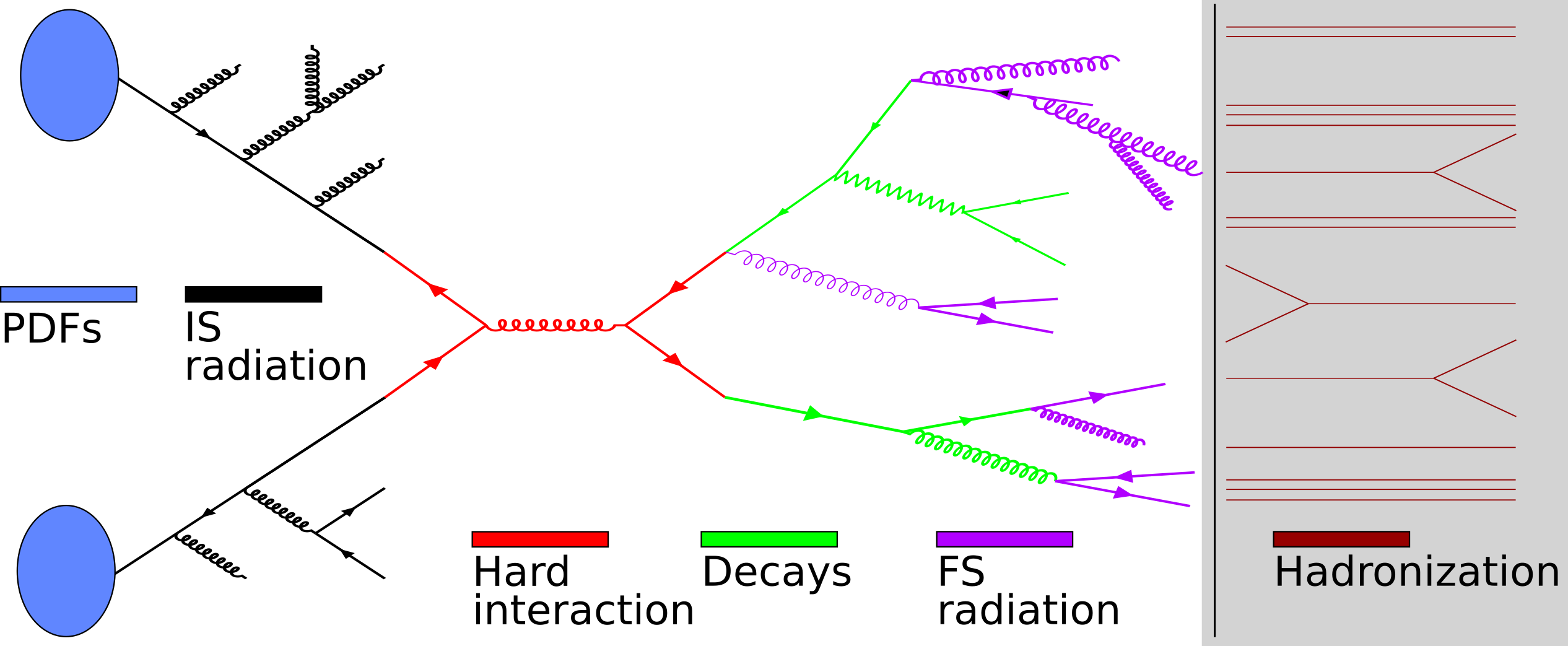}
\caption{The general layout of a hadron-hadron collision.}
\label{fig:overview}
\end{figure}

Monte Carlo event generators provide an essential tool for theoretical studies, phenomenological approaches, detector validations and for the planning of future facilities as well. They can combine several models together, and therefore it is possible to test the various aspects of high-energy collisions. The schematic structure of a proton-proton collision is shown on Figure~\ref{fig:overview}. However, we note, the \textit{soft physics} is neglected in the current study, since the factorization theorem is well-used in our approach to separate soft from hard processes. 

The theoretical description of hadronization (the main subject of the current study, denoted as the shaded area on Figure~\ref{fig:overview}) with the traditional perturbative approaches is a highly nontrivial task, and it is necessary to make physically motivated assumptions that can be tuned with experimental input. There is a variety of hadronization models in the literature --- the Lund string fragmentation is one of the most successful existing model, applied in the widely used {\sc Pythia} general purpose event generator~\cite{Sjostrand:1982fn, Sjostrand:2014zea}. By applying the appropriate set of tuned parameters, {\sc Pythia} is able to reproduce a wide range of experimentally measured data with a high precision --- however, the true nature of hadronization itself still remains to be studied.

Since it is such a crucial part of each event generator code, it is worthwhile to briefly investigate the computational cost of hadronization. By measuring the runtime fraction of a common configuration ({\sc Pythia} v8.3, proton-proton collisions at $\sqrt{s}=7$~TeV center-of-mass energy with standard Monash tune~\cite{Skands:2014pea}), it turns out that in a single event, 49\% of the time is spent solely with the fragmentation process (this ratio is approximately 9\% for proton-lead and 1\% for lead-lead collisions, utilizing the Angantyr model~\cite{Bierlich:2018xfw}). Considering the necessary amount of MC calculations (especially in the forthcoming HL-LHC era), this is not a negligible computational time (and thereby electricity cost), therefore the development of a hardware accelerated, Machine Learning supported method (even for more specialized cases) is well motivated.

\section{Final-state multiplicity as experimental observable}
\label{sec:observables}

The global event observables, based on the measured hadronic final-states provide rich information on the QCD processes of the collision event. In this study, the total number of charged hadrons (multiplicity) in a wide rapidity acceptance, $N_{ch, |\eta|<4}$ has been investigated in an event-by-event basis. We will test our model at various LHC center-of-mass-energy values, where the $\sqrt{s}$ energy scaling can be evaluated.

\section{Neural network design}
\label{sec:architecture}

ResNet architectures are convolutional neural networks (CNN) employed to overcome the vanishing gradient problem in deep neural networks, which can hinder training by causing early layers to receive minimal updates~\cite{279181}. By introducing identity mappings between layers, ResNets mitigate this issue, allowing the network to maintain efficient training even as it becomes deeper and more complex~\cite{10.1162/neco.1997.9.8.1735, he2015deep}. This enables more effective feature extraction from input data, such as images or video streams, leading to better performance in tasks like object recognition. Due to their relative simplicity and well known behavior, in our study ResNet networks with various parameter numbers (complexities) are applied to mimic the phase space evolution of hadron collisions.

\begin{figure}[h]
\centering
\includegraphics[width=0.65\linewidth]{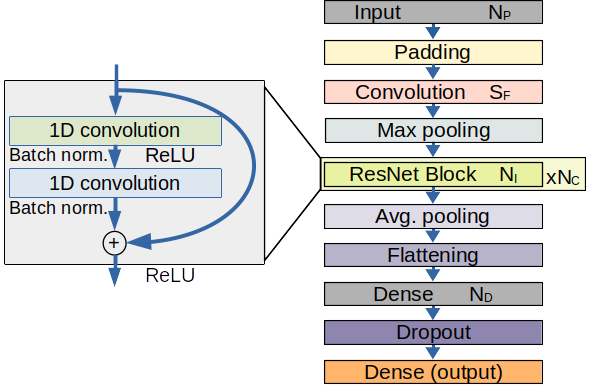}
\caption{The basic residual building block and the general structure of the applied ResNet models.}
\label{fig:layout}
\end{figure}

A basic building element is the \textit{residual block} with the identity mapping, which is sketched along with the general structure of the applied networks in Fig.~\ref{fig:layout}. The extent of complexity is determined by the number of input particles (see the following section in detail), $N_P$, the size of the convolutional filter bank, $S_F$, the number of convolutional and identity blocks in the ResNet structure, $N_C$ and $N_I$ respectively, and the number of dense nodes in the hidden layer, $N_D$. Generally, as it has been shown recently at the large language models~\cite{openai2023gpt4, devlin2019bert, touvron2023llama}, a deeper neural network with a larger number of trainable parameters ($N_{par}$) is able to adopt a better generalization of the trained model, and consequently have a superior performance. On the other hand, an important question of the NN design is that, what is the minimal complexity that is necessary to achieve the desired accuracy---in other words, how many trainable parameters are needed to describe the underlying physics. Therefore, focusing on the smaller parameter numbers, three different complexities are investigated in this study, summarized in Table~\ref{tab:models}.
\begin{table}[h]
  \centering
  \tbl{The configurations for the applied architectures.}{
  \begin{tabular}{lrrrrrr}
    \toprule
      Model & $N_P$ & $S_F$ & $N_C$ & $N_I$ & $N_D$ & $N_{par}$\\
    \colrule
    Small & 384 & 8 & 1 & 2 & 12 & 2,683 \\
    Medium & 384 & 16 & 2 & (3,3) & 16 & 28,179 \\
    Large & 384 & 32 & 3 & (3,4,3) & 32 & 416,291 \\
    \botrule
  \end{tabular}}
  \label{tab:models}
  \end{table}

At this point, an important remark must be made about determinism. Hadronization (and practically any physical process based on quantum mechanics) is a stochastic process, which means that for a given parton level input, the yield of the final state particles will follow some probability distributions. On the other hand, the proposed models are deterministic in the sense that there is no built-in sampling layer in them, which could produce stochastic behaviour. The purpose of the proposed hadronization models is to predict the event-by-event features at a macroscopic scale --- a more microscopic model with particle-by-particle capabilities is out of the scope of the current study. However, two completely identical phase-space configurations for two events are very improbable, and nearby configurations may mimic the probability distribution.

\begin{figure}[h]
  \centering
  \includegraphics[width=0.70\linewidth]{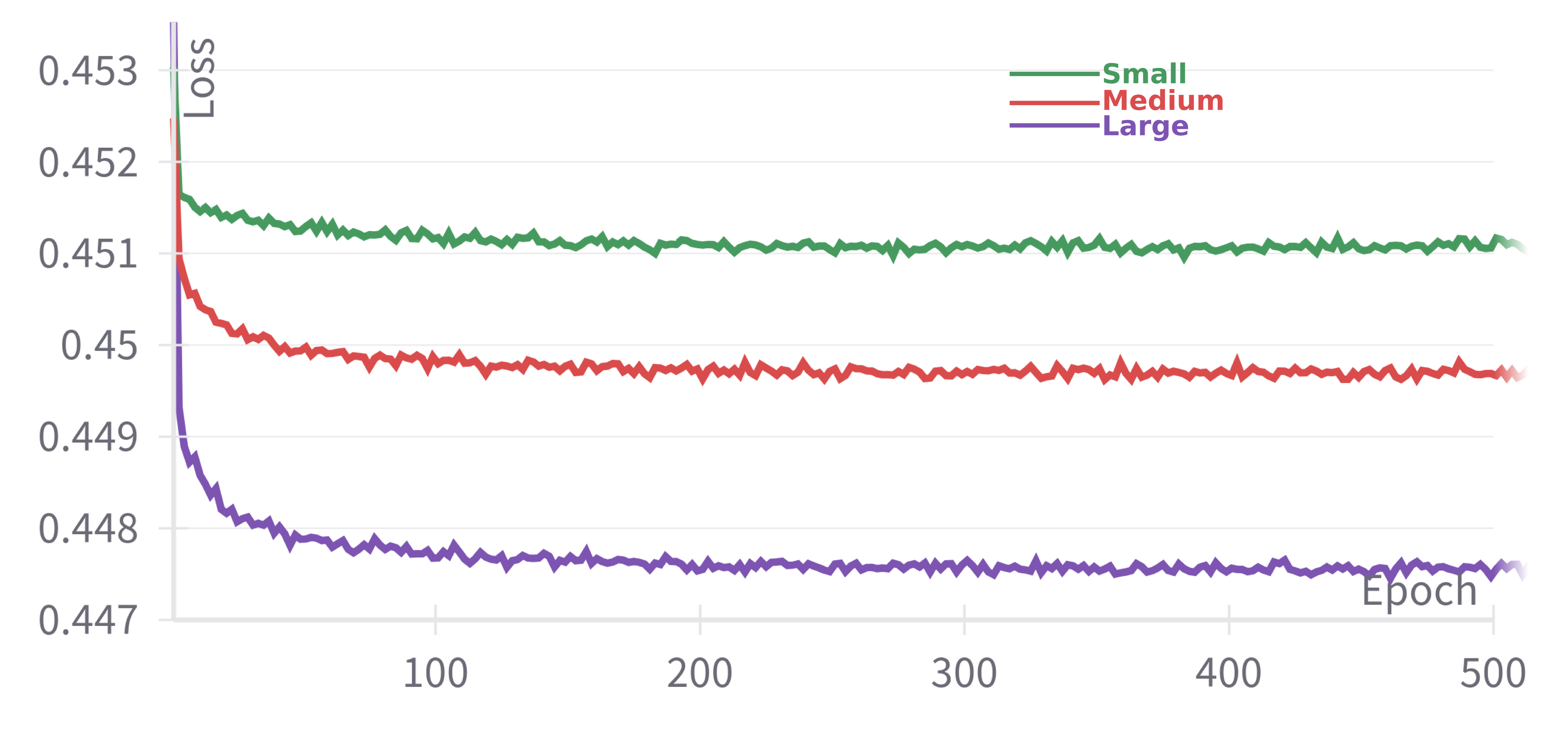}
  \caption{The loss value of the trained networks during 500 training epochs.}
  \label{fig:training}%
\end{figure}

In conventional applications, the main objective of these types of neural network architectures is usually some classification of the input images. However, in our implementation a set of physical quantities are inferred, that are observable in high-energy physics experiments. The output values are concatenated into one single vector, scaled into the $[0,1]$ region, therefore the sigmoid function has been chosen for the last activation function. The binary cross entropy is utilized as the loss function to approximate the event-by-event output as presented in Figure~\ref{fig:training}, while the Adam algorithm with default settings takes care for the optimization~\cite{kingma2017adam}. The initial learning rate of $0.01$ has been slowly decreased with a linear decay during 500 training epochs. 

Our models are implemented in Python, using Keras v2.7 with Tensorflow v.2.7 backend~\cite{chollet2015keras,abadi2016tensorflow}. The training, evaluating and testing were performed on a set of four Nvidia Tesla T4 graphics processing units of the Wigner Scientific Computing Laboratory (WSCLAB) at the HUN-REN Wigner Research Centre for Physics.

\section{Monte Carlo training data}
\label{sec:trainingdata}

Neither partonic degrees of freedoms, nor the hadronization process are directly (experimentally) observable; therefore, simulated events are needed to train and test the neural networks. For this purpose, the widely used {\sc Pythia} v8.3 general purpose event generator has been utilized with the commonly adopted Monash tune, that is known to reproduce LHC data with good accuracy~\cite{Sjostrand:1982fn, Sjostrand:2014zea, Skands:2014pea}. One of the main features of {\sc Pythia} is its string fragmentation model to perform the par\-ton-hadron transition~\cite{Sjostrand:1982fn}. The trained models are expected to generalize the main characteristics of this specific model. However, by future disentanglement of such a 'hadronizer' model might provide opportunities to study the unknown aspects of the hadronization process itself~\cite{Bellagente:2020piv}.

The input of the proposed neural networks is a two dimensional matrix, $M_{N_P, 4}$, where the rows contain the pseudorapidity, azimuth angle, the logarithm of the transverse momentum and the mass of each of the 'final-state' partons (i.e. the partons that are the input of the fragmentation process), scaled into the $[0, 1]$ region. Since the number of these partons varies from event to event, the rows of the zero-padded matrix is shuffled for each event. The output is a scalar containing the total multiplicity, scaled into the $[0, 1]$ region as well, determined by all final-state charged hadrons in the $|\eta|<4.0$ region of the given event.
\begin{figure}[h]
  \centering
  \includegraphics[width=0.60\linewidth]{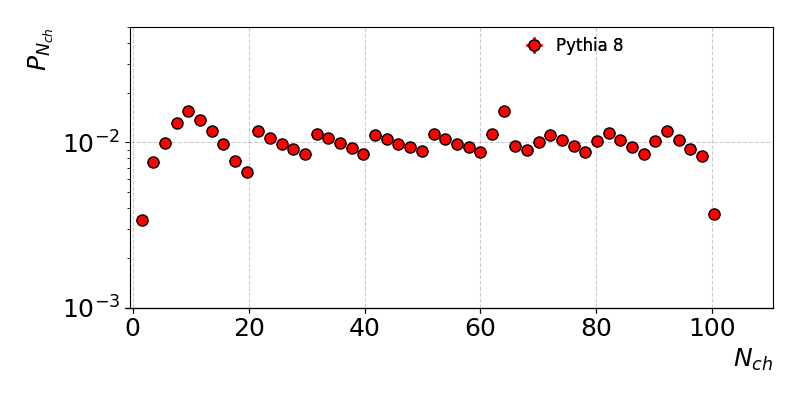}
  \caption{The uniformly-sampled multiplicity distribution of the training events.}
  \label{fig:training_mult}%
  \end{figure}
  
For the training and validation dataset, proton-proton collisions with a center-of-mass energy $\sqrt{s}=7$~TeV were quasi-uniformly-sampled with  multiplicity distribution: since the number of particles is one of the most significant features in the dataset, its distribution has to cover possible values more-or-less uniformly. In order to achieve this, each batch was required to contain an equal number of events within a properly chosen small range of multiplicity --- the multiplicity distribution of the training dataset is illustrated on Figure~\ref{fig:training_mult}.

For the test predictions, $\sqrt{s}=0.9$~TeV, $5.02$~TeV, and~$13$~TeV were considered as well, with regular minimum bias multiplicity distributions. For the training, $10^6$ events were generated, while the tests and validations were performed with an additional set of $10^5$ events.

It is important to recall that the event evolution in the {\sc Pythia} workflow consists of three main stages:
\begin{enumerate}
  \item Process Level, that includes the soft and hard QCD $2\rightarrow 2$ processes;
  \item Parton Level, including the initial- and final-state radiation, and the multiparton interactions (ISR, FSR and MPI, respectively);
  \item Hadron Level, incorporating the hadronization itself, with additional decays and rescatters.
\end{enumerate}
In order to ensure that it is really the hadronization process that is in the focus of the neural networks' training, the hadron level decays and rescatters have been turned off. 
Additionally, the initial assumption is that in the parametrization detailed above, the multiplicity distribution carries the main contribution of information---the main argumentation is that the hadronization is a well separated process in the {\sc Pythia} machinery, the parton-level processes occur independently. Though the ISR, FSR and MPI processes influence the parton-level multiplicity, they do not affect the forthcoming string fragmentation. Therefore, during the training we only used data where these partonic processes were turned on. We note, the trained hadronization networks were tested also with a dataset, where the ISR, FSR and MPI have been switched off, i.e. with events that solely contained the QCD $2\rightarrow 2$ processes. 

\section{Results}
\label{sec:results}

\begin{figure}[h]
    \centering
    \includegraphics[width=0.38\linewidth]{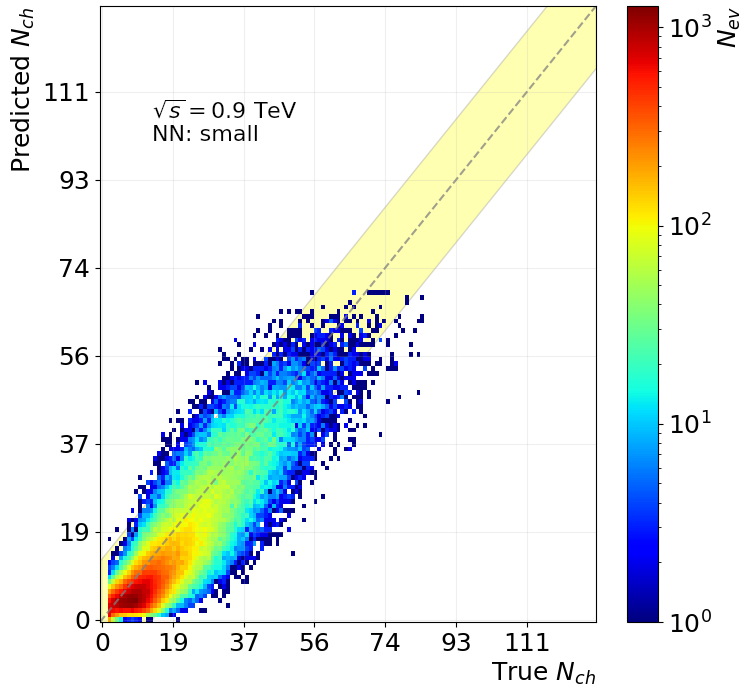} \hspace{-1.1truecm}
    \includegraphics[width=0.38\linewidth]{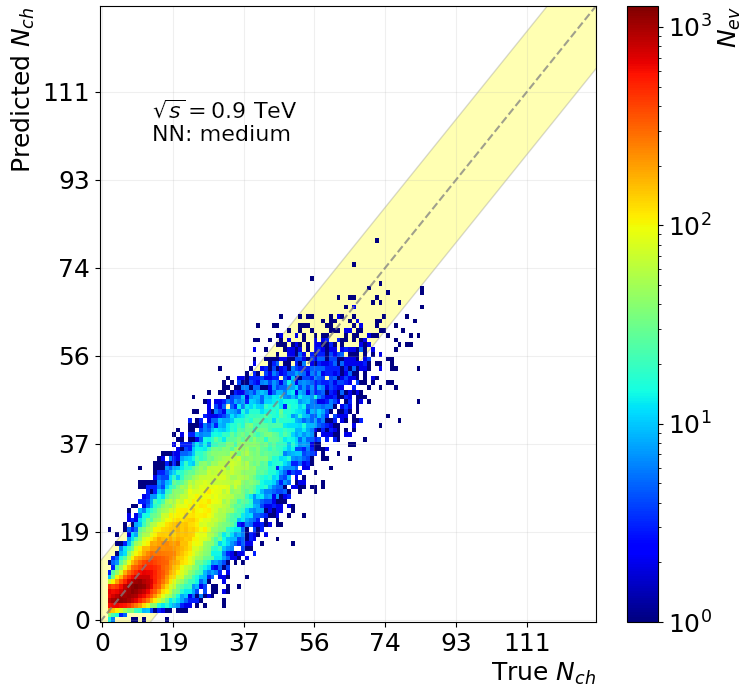} \hspace{-1.1truecm}
     \includegraphics[width=0.38\linewidth]{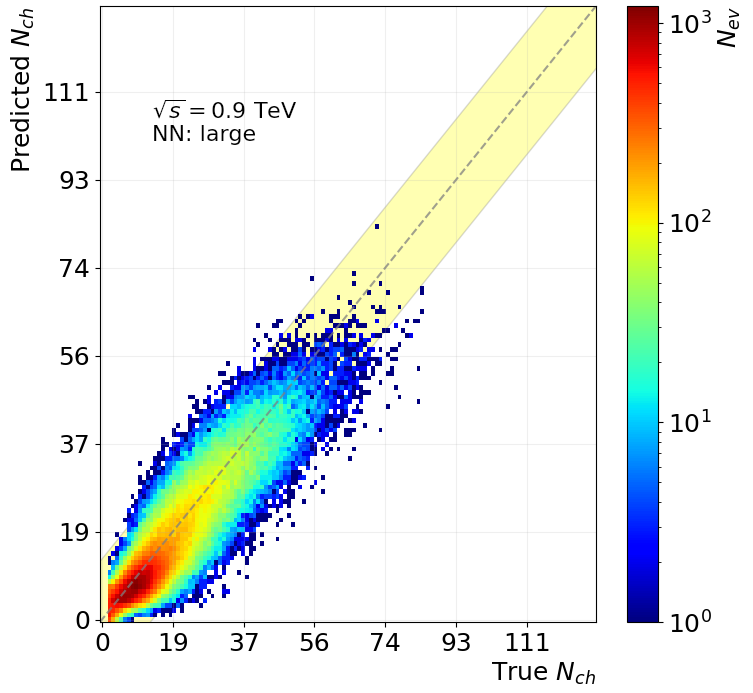}
     
    \includegraphics[width=0.38\linewidth]{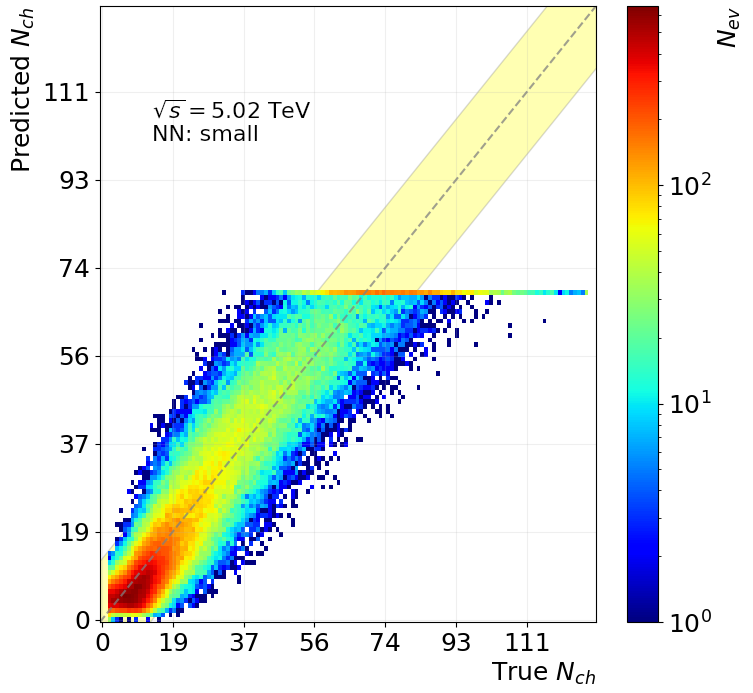} \hspace{-1.1truecm}
    \includegraphics[width=0.38\linewidth]{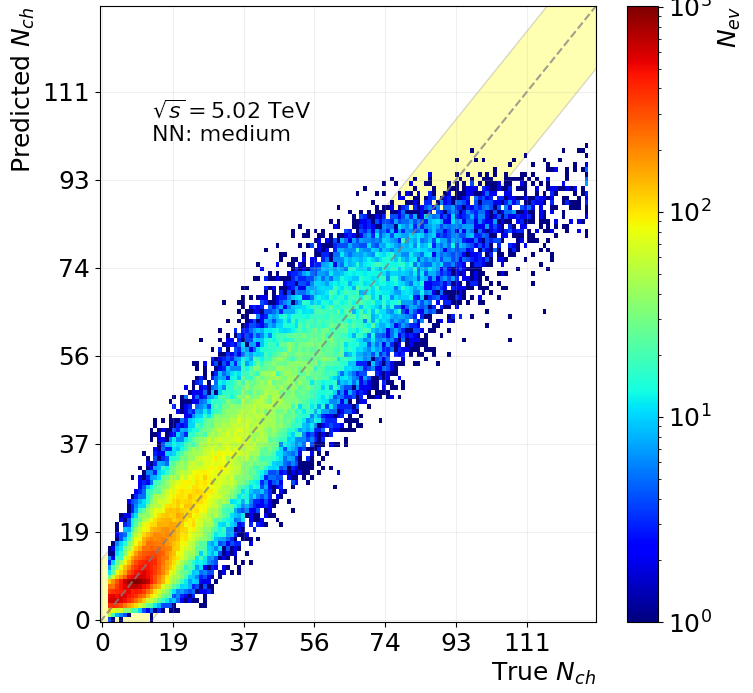} \hspace{-1.1truecm}
    \includegraphics[width=0.38\linewidth]{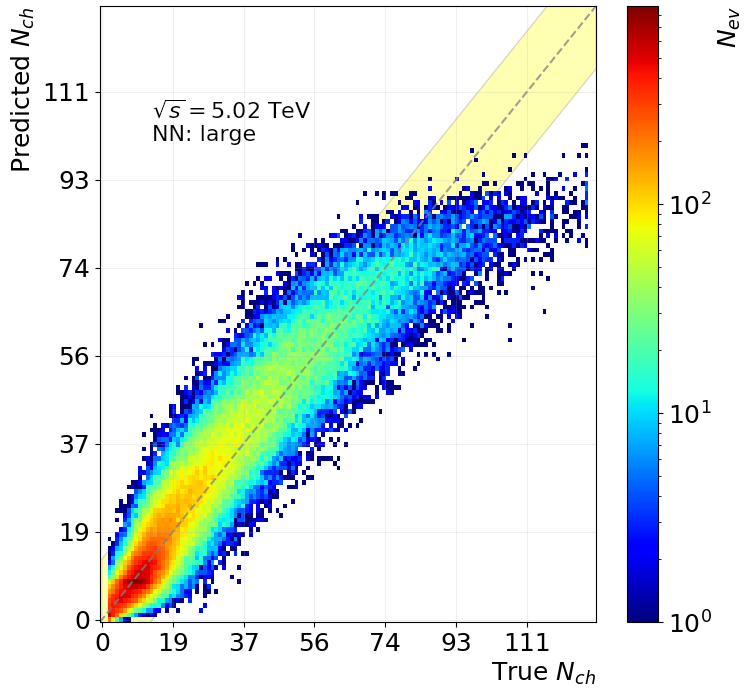}
     
    \includegraphics[width=0.38\linewidth]{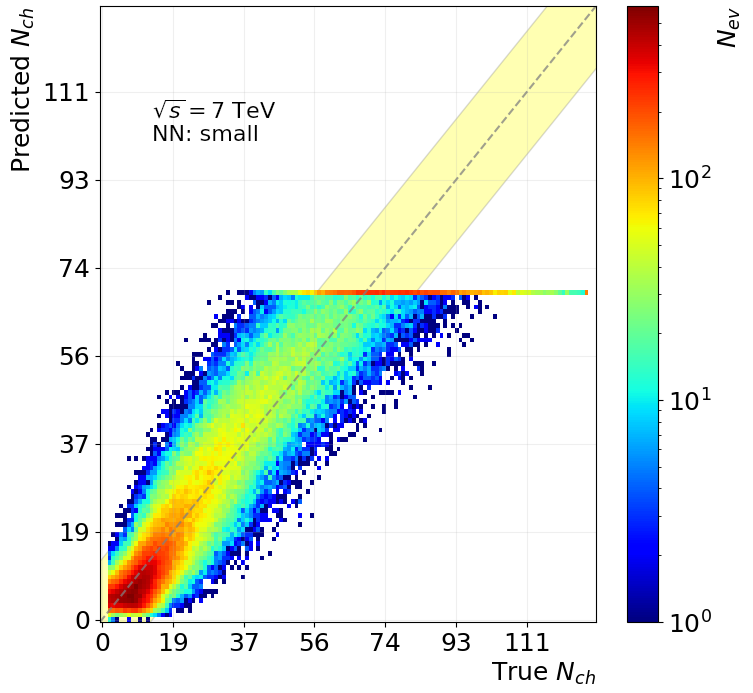} \hspace{-1.1truecm} 
     \includegraphics[width=0.38\linewidth]{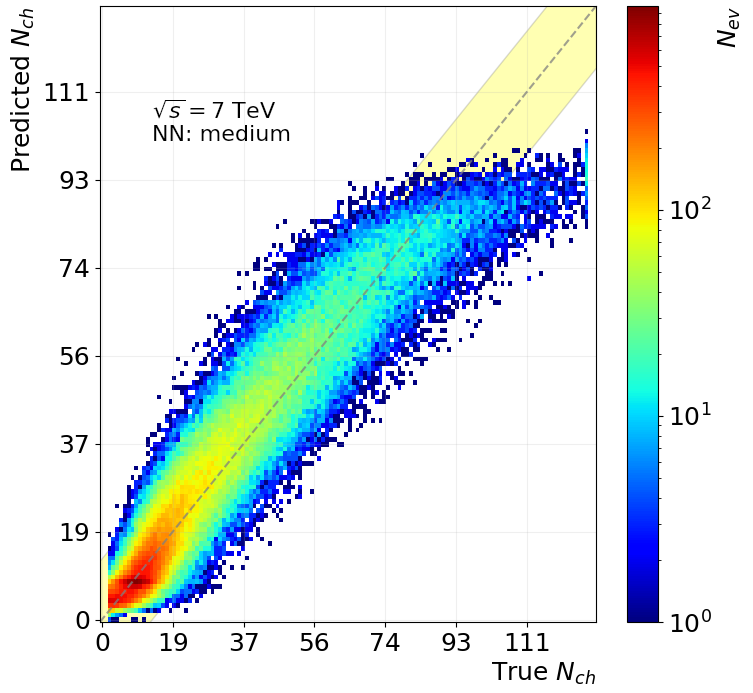} \hspace{-1.1truecm} 
     \includegraphics[width=0.38\linewidth]{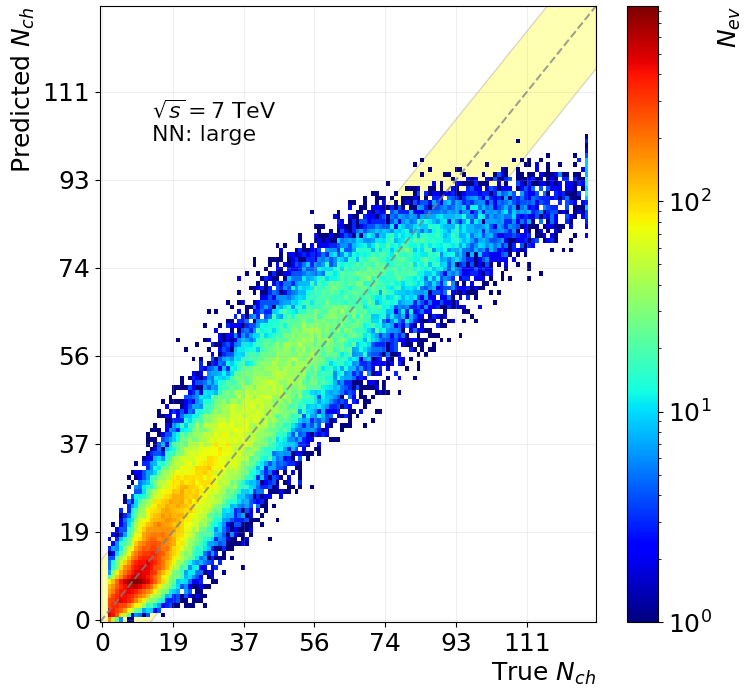} 
   
    \includegraphics[width=0.38\linewidth]{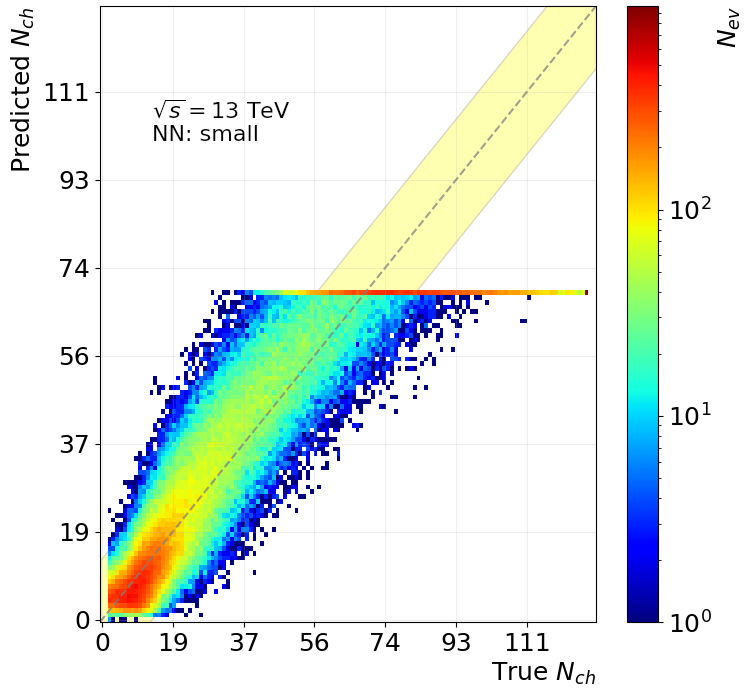} \hspace{-1.1truecm}
    \includegraphics[width=0.38\linewidth]{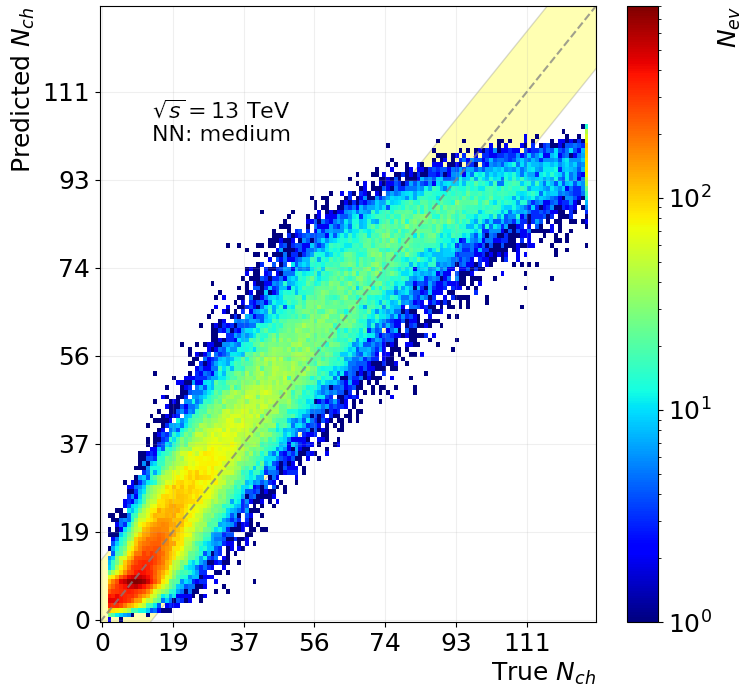} \hspace{-1.1truecm}
    \includegraphics[width=0.38\linewidth]{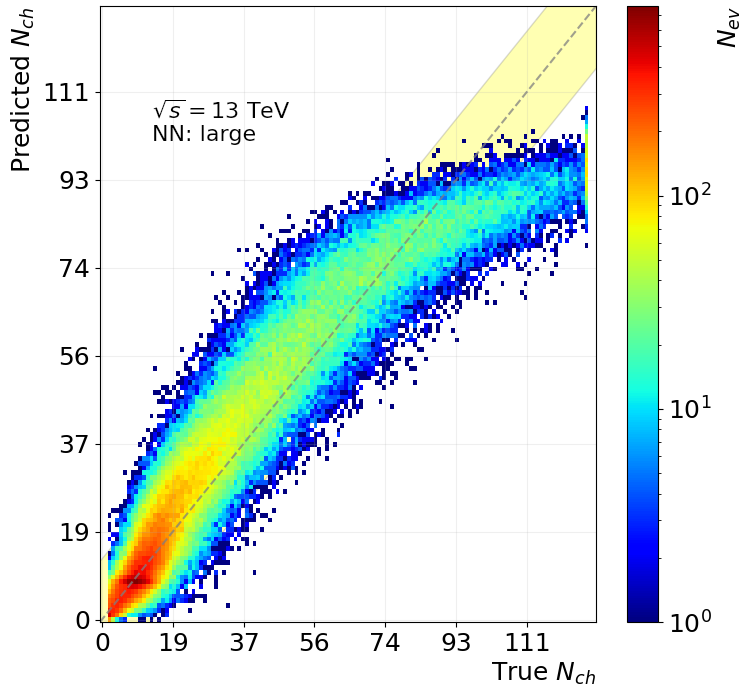} 
     \caption{The correlation plots of final-state charged hadrons at LHC energies with various NN complexities}
  \label{fig:mainresults}%
\end{figure}

Final state, unidentified charged hadron multiplicities ($N_\mathrm{ch}$) are shown for LHC energies generated in proton-proton collision in Figure~\ref{fig:mainresults}. We plotted the "Predicted" $N_\mathrm{ch}$ with respect to the Monte-Carlo generated "True" $N_\mathrm{ch}$ multiplicity values. 
We applied the 3 sets of neural network architectures presented in Table~\ref{tab:models}, and denoted as "Small", "Medium" and "Large" in columns from left to right, respectively. Rows represent certain energy values: 0.9~TeV, 5.02~TeV, 7~TeV and 13~TeV, from top to bottom. The 1-$\sigma$ band is presented by a yellow band on each panel.

Although, architectures are similar, the number of parameters can affect the correlations, especially in the high-multiplicity regime. Here, the minimal number of parameters limits the phase-phase, therefore the minimal-design of the "Small" model is unable to reach beyond $N_\mathrm{ch} \approx 70$ (see first columns). It is interesting to see, that architecture, with more than one order of magnitude higher number of parameters can predict well, up to the multiplicities $N_\mathrm{ch}\lesssim 90$ available with the 7~TeV training data. In architectures "Medium" and "Large" the correlation is getting off to identical and the predicted $N_\mathrm{ch}$ values are limited since predicted data was not constrained.

At the level of the development, we can sate, that this type of networks with more parameters than the "Small" version (e.g. $\gtrsim\mathcal{O}(10^3)$), the event-by-event predictive power has been achieved for multiplicity values $N_\mathrm{ch}\lesssim 90$. Comparing the multiplicity-scaling to the CNN model prediction, seemingly the upper multiplicity limit can be extended, by higher energy trainings.

\section{Conclusions}
\label{sec:summary}

In this study, popular convolutional neural network architectures with different number of parameters ranging from few to hundreds of thousands have been investigated in the context of hadronization of partonic states in high-energy hadron collisions, with different complexities. The training of the models were performed using simulated proton-proton events by the {\sc Pythia 8} Monte Carlo event generator at 7~TeV while the validation has been carried out including other LHC energies as well.

The simple models studied in this work were able to adopt the main features of the Lund string fragmentation at event-by-event basis. Moreover, the same trained models have been applied successfully to other center-of-mass energies, implying that the models could embrace the concept of multiplicity and energy scaling. It was also observed that learning of the produced multiplicity in hadronization works well if the networks sample the high-statistics part of the phase-space and the model parameters are greater than $\mathcal{O}(10^3)$. Seemingly this is a powerful feature, that can lead to various further research directions: a neural network with strong generalization capabilities could provide valuable input for the development and tuning process of Monte Carlo event generators. 

On the other hand, by reverse engineering the network, the properties of the process of hadronization might be investigated. In case of high-energy heavy-ion collisions, the same models might provide useful theoretical assist in the study of the parton-medium interactions as well.

\vspace{6pt} 


\section*{Acknowledgments}
This work has been supported by the NKFIH grants OTKA K135515, as well as by the 2021-4.1.2-NEMZ\_KI-2024-00031 and 2021-4.1.2-NEMZ\_KI-2024-00033 projects. The authors acknowledge the research infrastructure provided by the Hungarian Research Network (HUN-REN) and the Wigner Scientific Computing Laboratory. The authors are grateful to Antal Jakovác for the useful discussions.

Authors G. B. and G. P. were supported by the European Union project RRF-2.3.1-21-2022-00004 within the framework of the Artificial Intelligence National Laboratory.

\bibliographystyle{ws-ijmpa}
\bibliography{papers_all}

\end{document}